\documentclass[preprint,12pt,authoryear]{elsarticle}

\usepackage{kurz}
\usepackage{amssymb}
\usepackage{mathrsfs}
\usepackage{multirow, hhline}    
\usepackage[table,xcdraw]{xcolor}
\usepackage{amsmath}
\allowdisplaybreaks[4]
\usepackage{caption}
\usepackage{subcaption}
\usepackage{a4wide}
\usepackage{hyperref}
\hypersetup{hidelinks,
	colorlinks=true,
	allcolors=black,
	pdfstartview=Fit,
	breaklinks=true}
\usepackage{algorithm}
\usepackage{algpseudocode}
\usepackage{enumitem}

\usepackage{lineno}

\journal{arXiv}

\begin{document}
	
\begin{frontmatter}



\title{A multi-fidelity deep operator network (DeepONet) for fusing simulation and monitoring data: Application to real-time settlement prediction during tunnel construction}


\author[ISM-BO]{Chen Xu}
\author[ISM-BO]{Ba Trung Cao\corref{cor}\fnref{fncor}}
\author[HPC]{Yong Yuan}
\author[ISM-BO]{Günther Meschke}

\cortext[cor]{Corresponding author}
\fntext[fncor]{email: ba.cao@rub.de}

\address[ISM-BO]{Institute for Structural Mechanics, Ruhr University Bochum, Universitätsstraße 150, 44801 Bochum, Germany}
\address[HPC]{Department of Geotechnical Engineering, College of Civil Engineering, Tongji University, 1239 Siping Road, 20092 Shanghai, China}

\begin{abstract}
Ground settlement prediction during the process of mechanized tunneling is of paramount importance and remains a challenging research topic. Typically, two paradigms are existing:  a physics-driven approach utilizing process-oriented computational simulation models for the  tunnel-soil interaction and the settlement prediction, and a data-driven approach employing machine learning techniques to establish mappings between influencing factors and the ground settlement. To integrate the advantages of both approaches and to assimilate the data from different sources, we propose a multi-fidelity deep operator network (DeepONet) framework, leveraging the recently developed operator learning methods. The presented framework comprises of two components: a low-fidelity subnet that captures the fundamental ground settlement patterns obtained from finite element simulations, and a high-fidelity subnet that learns the nonlinear correlation between numerical models and real engineering monitoring data. A pre-processing strategy for causality is adopted to consider the spatio-temporal characteristics of the settlement during tunnel excavation. Transfer learning is utilized to reduce the training cost for the low-fidelity subnet. The results show that the proposed method can effectively capture the physical information provided by the numerical simulations and accurately fit measured data as well. Remarkably, even with very limited noisy monitoring data, the proposed model can achieve rapid, accurate, and robust predictions of the full-field ground settlement in real-time during mechanized tunnel excavation.
\end{abstract}

\begin{keyword}
DeepONet
\sep Multi-fidelity
\sep Data fusion
\sep Tunnel boring machine
\sep Ground settlement prediction
\sep Field reconstruction
\end{keyword}

\end{frontmatter}




\section{Introduction}
\label{sec:Intro}
The growing population, urbanization, and demand for high-speed mobility have necessitated the construction of underground transportation systems in the modern era. In recent years, a large number of tunnels has been constructed to improve existing transportation systems. Mechanized tunneling is a highly automated construction process, which has been widely applied to underground engineering projects under various geological and hydrological conditions \citep{Maidl_et_al:12}. However, the surface settlement induced by the tunnel boring machine (TBM) excavation can pose a threat to the stability of existing buildings above ground, making the real-time prediction of the expected ground settlements ahead of the tunnel face during tunneling a critical engineering issue.

Generally, in a tunnel project, a limited number of monitoring locations are predefined in the design phase, at which the temporal evolution of the ground settlements are recorded during the tunnel construction.
However, during the actual construction process, in particular in urban environments, it is desirable to predict the settlements induced by the tunnel advancement expected for the forthcoming excavation steps in order to allow adequately controlling the face and grouting pressure to keep the ground movements within prescribed limits \citep{Cao_Freitag_Meschke:16}. This information is needed for the complete field in the vicinity of the tunnel boring machine and not only in specific monitoring points.
Therefore, to better support the construction process, it is necessary to develop an efficient prediction model, which is capable to predict the complete settlement field at multiple surface points in real-time during the tunneling process.

With the rapid development of deep learning \citep{Lecun_2015_deeplearning}, scientific machine learning (SciML) has gained increasing attention across diverse disciplines in science and engineering. Supported by sufficient data, deep neural networks can learn functions between arbitrary inputs and outputs, enabling emulation and analysis of various complex physical systems \citep{JIANG2017394,JIANG2021107702,cabrera2023fusion,Zhang_2020_EvoNN,HE2022109125,HE2024121160,gu2023digital}. However, purely data-driven approaches heavily rely on both the quantity and quality of data. In most cases, available monitoring data in engineering projects are extremely limited, and such a small data set with great noise can lead to poor generalization of neural networks. To address this ubiquitous issue in systems with some physics and some data \citep{Karniadakis_2021_PIML}, a prevalent class of methods are proposed in recent years, known as physics-informed neural networks (PINNs) \citep{Raissi_2019_PINN,Xu2023,CHEW2022132455,ZOBEIRY2021104232,FERNANDEZ2023105790,HAGHIGHAT2023105828,ZHANG2023106073,DING2023106425,SUN2023106742}. By leveraging the automatic differentiation technique, physical laws in the form of partial differential equations (PDEs) can be easily embedded into the loss function, thus reducing the requirements for data volume and improving the generalization performance of neural networks. However, the spatio-temporal ground settlement induced by TBM excavation is very complex, which can not be fully described via analytical solutions. In other words, the lack of physics to be incorporated into neural networks greatly limits the practical application of PINNs in this field.

More recently, the neural operator method has emerged as a new and promising research field in the SciML community, with notable methods such as the Deep Operator Network (DeepONet) proposed by Lu et al. \citep{LuLu_2021_DeepONet} and the Fourier neural operator (FNO) proposed by Li et al. \citep{ZongyiLi_2020_FNO}. Unlike previous approaches, the neural operator can learn mappings between infinite-dimensional Banach spaces \citep{Lanthaler_2022_DONerror,Kovachki_2021_FNOerror}. Through various methodological improvements \citep{Goswami_2022_transferDeepONet,SifanWang_2021_PIDeepONet,Zhu_2022_DONextrapolate,Lu_2022_DONcompareFNO,GARG2023105685}, it has demonstrated remarkable effectiveness in a wide range of scientific problems \citep{Lin_2021_DONbubble,Cai_2021_DeepMMnet,Goswami_2022_fractureDeepONet,MinglangYin_2022_interfaceDeepONet,Pickering_2022_DONextreme,DiLeoni_2021_dataassimi,SifanWang_2023_longtimeDON,LIN2023106689}.

In our previous work \citep{Cao_Freitag_Meschke:16,Freitag_Cao_Ninic:18,Cao_Obel_Freitag:20,Zendaki_et_al:23a}, a simulation-based surrogate modelling strategy for predictions of settlement fields in real time during TBM construction of tunnels is presented. This surrogate model is completely trained on synthetic data obtained from numerical simulations (i.e. low-fidelity data). Consequently, the predicted displacement field reflects the physical trends but might deviate from actual monitoring data (i.e. high-fidelity data) without any calibration. To maintain the validity of the prediction model, new available monitoring data is continuously integrated into the model under a retraining strategy.

In this paper, we propose a framework for Multi-Fidelity DeepONet to enable the seamless fusion of information from the two types of data mentioned above. Since the tunneling process is time-dependent, the Causality-DeepONet \citep{Liu_2022_CausalDON} is adopted for data pre-processing. This framework consists of two parts: the low-fidelity DeepONet trained on simulation-based data to learn the underlying physical mechanism during tunnel excavations, and the residual DeepONet trained on very limited monitoring data to bridge the gap between numerical models and measurements. The final reconstructed settlement field can exhibit physical patterns similar to numerical simulation results and fit well with the monitoring data. 

To the best of our knowledge, no study has ever managed to use the operator learning technique for settlement prediction in tunnel construction. The proposed multi-fidelity DeepONet framework has the following characteristics:

\begin{itemize}	
	\item [$\bullet$]\textbf{Fast}: It is an efficient architecture that combines the characteristics of two recent advances in DeepONet \citep{Lu_2022_multi-fidelity,Howard_2022_fidelityDON}. Transfer learning is employed at the online stage to reduce the computational cost. The low-fidelity subnet serves as a pre-trained model, facilitating rapid settlement reconstruction within only one minute.
	\item [$\bullet$]\textbf{Flexible}: The settlement of arbitrary points on the ground surface can be predicted even their location information were not included in the training process.
	\item [$\bullet$]\textbf{Accurate}: The model achieves a favorable result with the $R^2$ score around 0.9 on the task of settlement reconstruction at the online learning stage.
	\item [$\bullet$]\textbf{Robust}: Our model showcases excellent resilience against noise perturbations (typically occurring during measurement), with $R^2$ scores above 0.8 even under scenarios characterized by very limited noisy data.
\end{itemize}

The paper is organized as follows: In section~\ref{sec:Method}, after introducing some preliminaries of the finite element model for data set generation, we provide a brief description of the improved DeepONet architecture and present a framework of the multi-fidelity DeepONet with causality. In section~\ref{sec:Resu}, we demonstrate the effectiveness of the proposed approach in addressing the ground settlement reconstruction in tunneling. The conclusion and future works are included in the section~\ref{sec:Conc}.

\section{Methods}
\label{sec:Method}
\subsection{Finite Element Simulation For Mechanized Tunneling}
\label{sec:FEM}
\begin{figure}[!t]
	\centering
	\includegraphics[width=\textwidth]{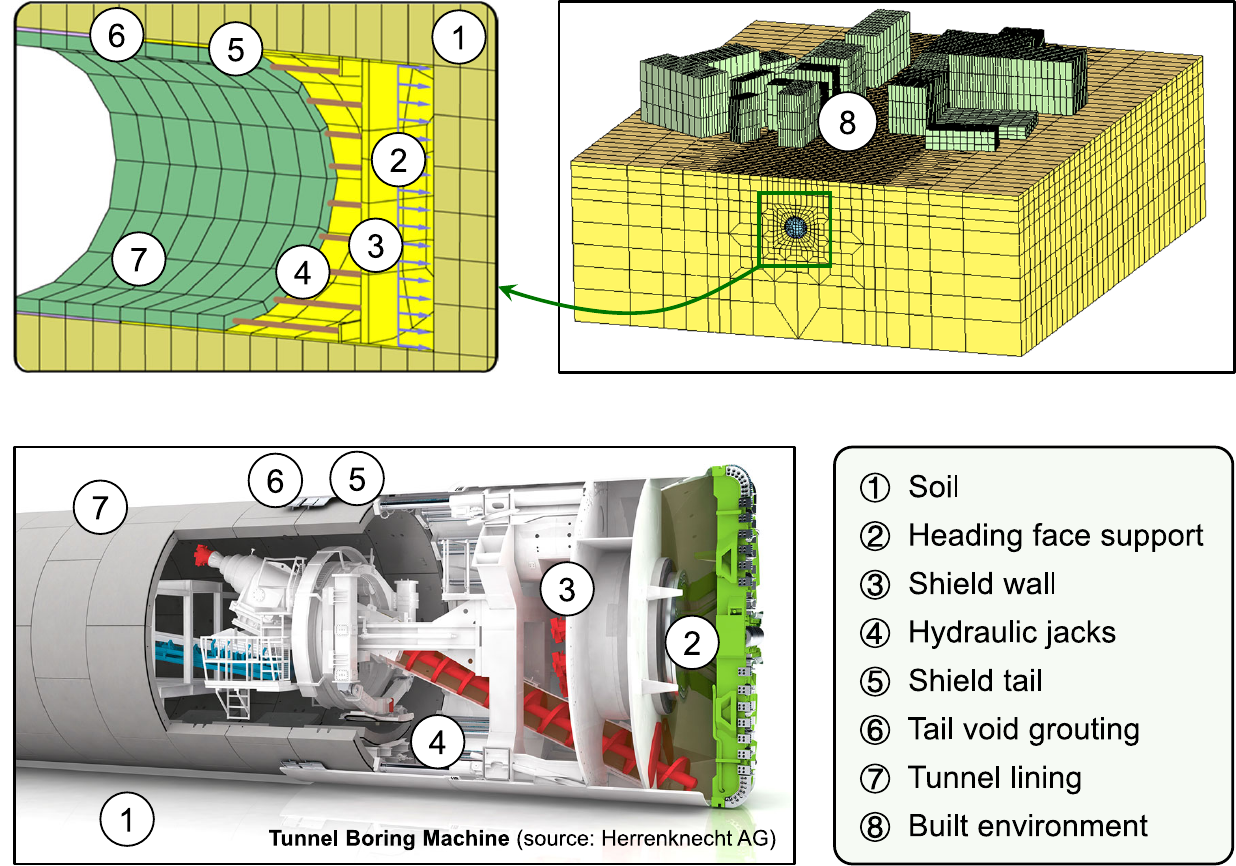}
	\caption{Components of the Finite Element Simulation model for tunnel excavation in soft soils.}
	\label{fig:fem}
\end{figure}

In this paper, an advanced 3D process-oriented Finite Element (FE) simulation model \citep{Alsahly_Stascheit_Meschke:16, Cao_Freitag_Meschke:16, Freitag_Cao_Ninic:18} is used to simulate mechanized tunneling processes in fully or partially saturated soft soils. 
Simulation results from the numerical model have been validated with data from tunnel projects such as the Wehrhahn metro line in Düsseldorf, Germany \citep{Ninic_Freitag_Meschke:17, Bui_Meschke:20}, which shows, in general, a good agreement between the simulation and measurement data.
The model, which takes into account all interactions between components of the tunnel construction process (see Fig.~\ref{fig:fem}), is therefore capable of representing the construction of tunnels using tunnel boring machines (TBM) in real-world problems.

In the simulation of a tunnel excavation employing a TBM, a step-by-step procedure following the actual tunnel advance and construction steps is carried out.
More specifically, in each TBM advance, soil elements in front of the TBM are deactivated to represent the excavation process. New lining ring elements and tail void grouting elements are activated to reflect the new situation at the back of the TBM, i.e. the installation of the tunnel lining and the filling of grout material into the annular gap between the lining and the excavation boundary.
Respective boundary conditions, e.g. the face support pressure $\boldsymbol{P_S}$ to ensure the stability of the tunnel face and the grouting pressure $\boldsymbol{P_G}$ to reduce ground loss behind the tapered shield, are also adapted on new and old surfaces accordingly.

Separated components are modeled to represent the main components of the construction process, such as the surrounding soil, the TBM, the tunnel lining, the tail void grouting, and the built environment on the ground surface.
Depending on the presence of air, water and soil solids, the surrounding soil can be modelled as a simple kinematic linear material (only solids), a two phase material (solids and water) in case of a fully saturated condition, and a three phase material (air, water and solids) in case of a partially saturated condition \citep{Cao_Freitag_Meschke:16}.
The inelasticity of soft soils can be simulated using two well-known elastoplastic models in geotechnics: the Clay and Sand model and the Drucker-Prager model.
To realistically simulate the deformation of the ground considering the real, tapered geometry, and the overcut of the TBM, the TBM is modeled as a 3D deformable body moving and interacting with the surrounding soil via frictional surface-to-surface contact.
In a realistic simulation approach, the movement of the TBM is simulated using a steering control algorithm based on the movement of individual hydraulic jacks \citep{Alsahly_Stascheit_Meschke:16}, which are modeled as truss elements and attached to the lining and the TBM wall. 
The tunnel lining can be modeled as a full ring with given length, diameter, and thickness, whereas the grouting material is characterized using a fully saturated two phase material with a hydrating matrix phase \citep{Meschke:96}.
In the simulation model, the built environment representing existing infrastructures on the ground surface such as buildings and transportation systems, can be modeled in different level of details (LoD).
In the lowest LoD, simple surrogate structures, e.g. 2D structures with equivalent stiffness and thickness using shell elements \citep{Cao_Obel_Freitag:20} or 3D blocks with a converted stiffness \citep{Cao_et_al:22}, can be adopted to model the actual structures.
Alternatively, for a detailed modeling expectation, sophisticated models with volume elements and structural connections \citep{Zendaki_et_al:23a}, 
can also be created and simulated as the highest LoD for the simulation. 
The interactions between the building foundation with the soil are considered through an interface mechanism based on the mortar tying formulation \citep{Bui_et_al:23}, which allows to simulate different mechanisms of the soil-structure interaction corresponding to so-called “sagging” and “hogging” modes.

\subsection{A Modified DeepONet Architecture}
\label{sec:ModiDON}
\begin{figure}[!t]
	\centering
	\includegraphics[scale=0.76]{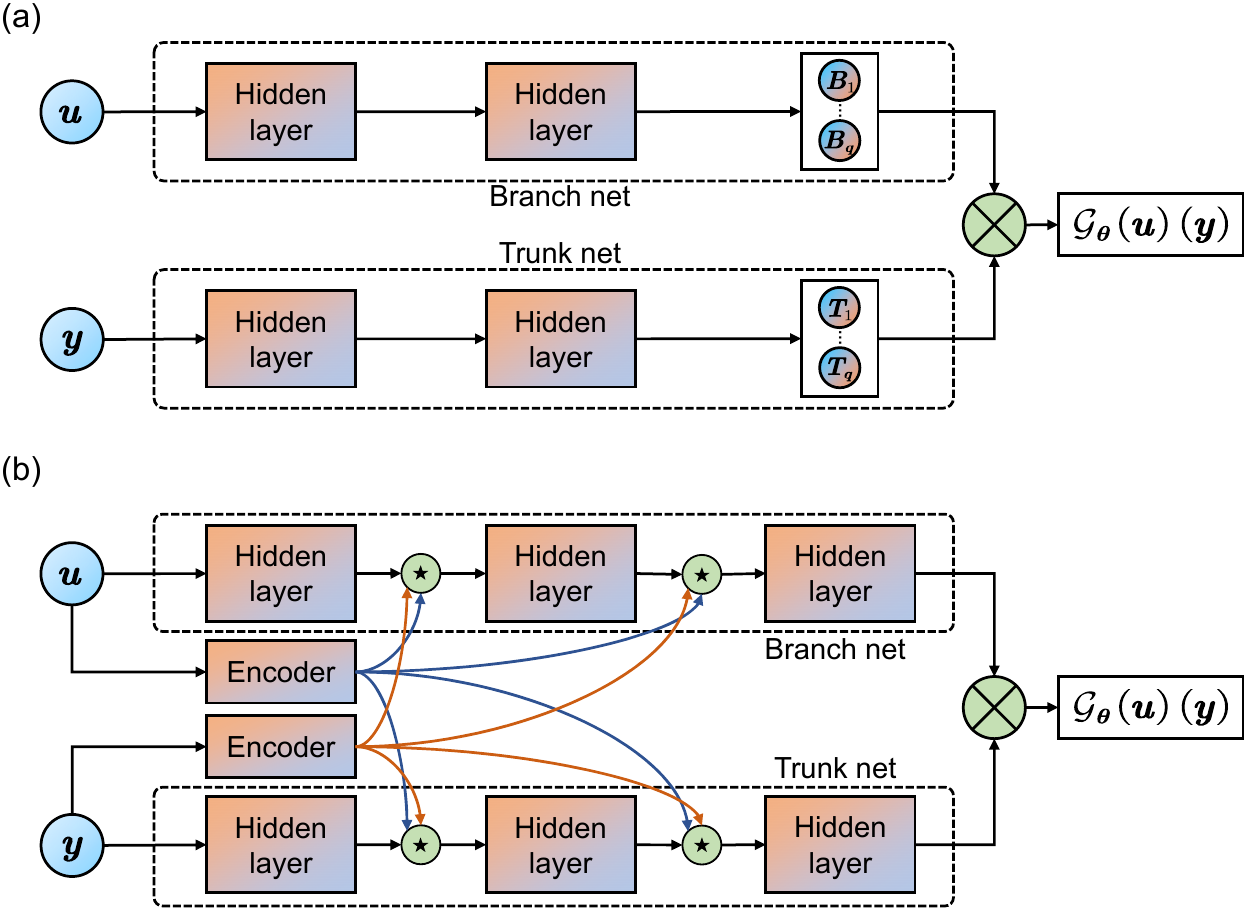}
	\caption{Schematic diagram of (a) the vanilla DeepONet and (b) the modified DeepONet architecture. The symbol $\bigstar$ denotes the operation defined in Eqs. \eqref{eq:star1} and \eqref{eq:star2}.}
	\label{fig:DeepONet}
\end{figure}

DeepONet was originally designed based on the universal approximation theorem for operators \citep{ChenChen_1995_uniappro}, aiming to learn nonlinear continuous operators between infinite-dimensional function spaces. Let $\mathcal{U}$ and $\mathcal{S}$ be the spaces of functions $\boldsymbol{u}$ and $\boldsymbol{s}$, respectively, and the objective is to learn the mapping operator $\mathcal{G}\,:\,\mathcal{U}\ni \boldsymbol{u}\,\mapsto \,\boldsymbol{s}\in \mathcal{S}$. A standard DeepONet consists of two neural networks: a branch net and a trunk net (see Fig.~\ref{fig:DeepONet}(a)). The branch net takes the function $\boldsymbol{u}$ evaluated at points $\left\{ r_1,r_2,\cdots ,r_m \right\}$ as input and extracts embedded features of the discrete input function space, yielding $\left[ B_1,B_2,\cdots ,B_q \right] ^T$ as output. The trunk net takes the coordinates $\boldsymbol{y}$ of function $\boldsymbol{s}$ as input and the output is $\left[ T_1,T_2,\cdots ,T_q \right] ^T$. The final DeepONet output is obtained by combining the outputs of branch and trunk nets through an inner product, which is expressed as:
\begin{equation}
\begin{aligned}
\mathcal{G}_{\boldsymbol{\theta }}\left( \boldsymbol{u} \right) \left( \boldsymbol{y} \right) =\sum_{k=1}^q{B_k\left( \boldsymbol{u}\left( \boldsymbol{r}_1 \right) ,\boldsymbol{u}\left( \boldsymbol{r}_2 \right) ,\cdots ,\boldsymbol{u}\left( \boldsymbol{r}_{\boldsymbol{m}} \right) \right)}T_k\left( \boldsymbol{y} \right) .
\end{aligned}
\end{equation}

In this work, a novel modified DeepONet architecture proposed in Ref.~\citep{Wang_2022_improvedDON} is adopted. As depicted in Fig.~\ref{fig:DeepONet}(b), two additional encoders are introduced in the modified DeepONet to further encode the input signals of the branch and trunk nets. These encoders can facilitate the propagation of coded information through point-wise multiplication in each hidden layer, thereby enhancing the nonlinear expressive capabilities of the neural network. The detailed forward process for an $L-$layer DeepONet is given as follows:
\begin{align}
&\boldsymbol{U}=\phi \left( \boldsymbol{W}_u\boldsymbol{u}+\boldsymbol{b}_u \right) ,\quad \boldsymbol{V}=\phi \left( \boldsymbol{W}_y\boldsymbol{y}+\boldsymbol{b}_y \right) 
\\
&\boldsymbol{H}_{u}^{(1)}=\phi \left( \boldsymbol{W}_{u}^{(1)}\boldsymbol{u}+\boldsymbol{b}_{u}^{(1)} \right) ,\quad \boldsymbol{H}_{y}^{(1)}=\phi \left( \boldsymbol{W}_{y}^{(1)}y+\boldsymbol{b}_{y}^{(1)} \right) 
\\
&\boldsymbol{Z}_{u}^{(l)}=\phi \left( \boldsymbol{W}_{u}^{(l)}\boldsymbol{H}_{u}^{(l)}+\boldsymbol{b}_{u}^{(l)} \right) ,\quad \boldsymbol{Z}_{y}^{(l)}=\phi \left( \boldsymbol{W}_{y}^{(l)}\boldsymbol{H}_{y}^{(l)}+\boldsymbol{b}_{y}^{(l)} \right) ,\quad l=1,2,...,L-1
\\
&\boldsymbol{H}_{u}^{(l+1)}=\left( 1-\boldsymbol{Z}_{u}^{(l)} \right) \odot \boldsymbol{U}+\boldsymbol{Z}_{u}^{(l)}\odot \boldsymbol{V},\quad l=1,...,L-1
\label{eq:star1}
\\
&\boldsymbol{H}_{y}^{(l+1)}=\left( 1-\boldsymbol{Z}_{y}^{(l)} \right) \odot \boldsymbol{U}+\boldsymbol{Z}_{y}^{(l)}\odot \boldsymbol{V},\quad l=1,...,L-1
\label{eq:star2}
\\
&\boldsymbol{H}_{u}^{(L)}=\phi \left( \boldsymbol{W}_{u}^{(L)}\boldsymbol{H}_{u}^{(L-1)}+\boldsymbol{b}_{u}^{(L)} \right) ,\quad \boldsymbol{H}_{y}^{(L)}=\phi \left( \boldsymbol{W}_{y}^{(L)}\boldsymbol{H}_{y}^{(L-1)}+\boldsymbol{b}_{y}^{(L)} \right) 
\\
&\mathcal{G}_{\boldsymbol{\theta}}(\boldsymbol{u})(\boldsymbol{y})= \langle \boldsymbol{H}_{u}^{(L)},\boldsymbol{H}_{y}^{(L)} \rangle ,
\end{align}
where $\boldsymbol{U}$ represents the branch encoder with weights $\boldsymbol{W}_u$ and biases $\boldsymbol{b}_u$, while $\boldsymbol{V}$ denotes the trunk encoder with weights $\boldsymbol{W}_y$ and biases $\boldsymbol{b}_y$. $\left\{ \boldsymbol{W}_{u}^{\left( l \right)},\boldsymbol{b}_{u}^{\left( l \right)} \right\} _{l=1}^{L}$ and $\left\{ \boldsymbol{W}_{y}^{\left( l \right)},\boldsymbol{b}_{y}^{\left( l \right)} \right\} _{l=1}^{L}$ are weights and biases of the branch net and trunk net, respectively. $\phi$ represents the activation function. $\odot$ denotes the Hadamard product, while $\langle\cdot,\cdot\rangle$ indicates the inner product.  $\boldsymbol{\theta}$ encompasses all trainable parameters within the entire neural network.

\subsection{A Multi-fidelity DeepONet With Causality}
Followed by the finite element simulations introduced in section~\ref{sec:FEM}, a large low-fidelity data set $\mathcal{T}_L=\left\{ \boldsymbol{u}_{L}^{i},\boldsymbol{y}_{L}^{i},\boldsymbol{s}_{L}^{i} \right\} _{i=1}^{N_L}$ can be generated. Here, $\boldsymbol{u}_{L}$ represents the TBM process function (i.e., tail void grouting pressure $\boldsymbol{P_G}$ and face support pressure $\boldsymbol{P_S}$ in this paper), $\boldsymbol{y}_{L}$ denotes the coordinates of points on the ground surface, and $\boldsymbol{s}_{L}$ is the tunneling-induced settlements located at $\boldsymbol{y}_{L}$. 

It is noted that the TBM process parameters $\boldsymbol{P_G}$ and $\boldsymbol{P_S}$ are time-variants, and the resulted ground settlement at the current step is not affected by the future TBM process, but only depends on its current state and past history. This temporal causality characteristic has to be taken into consideration. Consequently, we adopt the Causality-DeepONet \citep{Liu_2022_CausalDON} for pre-processing the data set. Specifically, for a specific scenario in $\mathcal{T}_L$, the low-fidelity data take the form of a triplet, $\left[ \left\{ \boldsymbol{u}_{L}^{ t_i } \right\} _{t_i=1}^{N_t}, \left\{ \boldsymbol{y}_{L}^{ j } \right\} _{j=1}^{N_{lp}}, \boldsymbol{s}_L\left( \boldsymbol{u}_L \right) \left( \boldsymbol{y}_L \right) \right]$:
\begin{equation}
\begin{aligned}
\left[ \begin{array}{c}
	\boldsymbol{u}_{L}^{ t_i }=\left[ \boldsymbol{P_G}\left( t_i \right) ,\boldsymbol{P_G}\left( t_{i-1} \right) ,\boldsymbol{P_G}\left( t_{i-2} \right) ,\cdots ,\boldsymbol{P_G}\left( t_2 \right) ,\boldsymbol{P_G}\left( t_1 \right) ,\overset{N_t-t_i}{\overbrace{0,\cdots ,0}} \right. ,\\
	\left. \boldsymbol{P_S}\left( t_i \right) ,\boldsymbol{P_S}\left( t_{i-1} \right) ,\boldsymbol{P_S}\left( t_{i-2} \right) \cdots ,\boldsymbol{P_S}\left( t_2 \right) ,\boldsymbol{P_S}\left( t_1 \right) ,\overset{N_t-t_i}{\overbrace{0,\cdots ,0}} \right] ,\\
	\boldsymbol{y}_{L}^{ j }=\left[ \boldsymbol{x}_{1}^{ j },\boldsymbol{x}_{2}^{ j } \right] ,\quad \quad \boldsymbol{s}_L\left( \boldsymbol{u}_{L}^{ t_i } \right) \left( \boldsymbol{y}_{L}^{ j } \right)\\
\end{array} \right], 
\end{aligned}
\label{eq:triplet}
\end{equation}
here, an individual scenario is composed of $N_t$ discrete time steps, with each time step requiring the records of ground settlement data for $N_{lp}$ specific points. If the complete data set $\mathcal{T}_L$ encompasses $N_s$ distinct scenarios, then the total number of data records $N_L$ within the data set $\mathcal{T}_L$ should be equal to $N_s \times N_t \times N_{lp}$. For a given scenario, the branch input $\boldsymbol{u}_{L}^{t_i}$ at the current time step $t_i$ is formed by concatenating two $N_t$-dimensional vectors, each of which is related to the TBM process parameters $\boldsymbol{P_G}$ and $\boldsymbol{P_S}$, respectively. Considering the intrinsic causal nature of $\boldsymbol{u}_{L}^{t_i}$, an additional convolution operation coupled with zero-padding has to be performed. Taking the first half of $\boldsymbol{u}_{L}^{t_i}$, which corresponds to the $\boldsymbol{P_G}$ series to be input, as an illustrative example, the first $t_i$ elements are arranged in a reverse chronological order, while the subsequent $N_t-t_i$ elements are all set to zero. This strategy can effectively prevent future $\boldsymbol{P_G}$ values from affecting ground settlement at the current time step.

As for the high-fidelity data set $\mathcal{T}_H=\left\{ \boldsymbol{u}_{H}^{i},\boldsymbol{y}_{H}^{i},\boldsymbol{s}_{H}^{i} \right\} _{i=1}^{N_H}$, it can be acquired through in-situ monitoring in engineering projects. The data set $\mathcal{T}_H$ is typically characterized by a much smaller data volume, i.e., $N_{H}\ll N_{L}$. The structure of the data set $\mathcal{T}_H$ remains consistent with that described in Eq. \eqref{eq:triplet}, which is also organized in the form of a triplet, $\left[ \left\{ \boldsymbol{u}_{H}^{t_i} \right\} _{t_i=1}^{t_n},\left\{ \boldsymbol{y}_{H}^{j} \right\} _{j=1}^{N_{hp}},\boldsymbol{s}_H\left( \boldsymbol{u}_H \right) \left( \boldsymbol{y}_H \right) \right]$:
\begin{equation}
\begin{aligned}
\left[ \begin{array}{c}
	\boldsymbol{u}_{H}^{ t_i }=\left[ \boldsymbol{P_G}\left( t_i \right) ,\boldsymbol{P_G}\left( t_{i-1} \right) ,\boldsymbol{P_G}\left( t_{i-2} \right) ,\cdots ,\boldsymbol{P_G}\left( t_2 \right) ,\boldsymbol{P_G}\left( t_1 \right) ,\overset{N_t-t_i}{\overbrace{0,\cdots ,0}} \right. ,\\
	\left. \boldsymbol{P_S}\left( t_i \right) ,\boldsymbol{P_S}\left( t_{i-1} \right) ,\boldsymbol{P_S}\left( t_{i-2} \right) \cdots ,\boldsymbol{P_S}\left( t_2 \right) ,\boldsymbol{P_S}\left( t_1 \right) ,\overset{N_t-t_i}{\overbrace{0,\cdots ,0}} \right] ,\\
	\boldsymbol{y}_{H}^{ j }=\left[ \boldsymbol{x}_{1}^{ j },\boldsymbol{x}_{2}^{ j } \right] ,\quad \quad \boldsymbol{s}_L\left( \boldsymbol{u}_{H}^{ t_i } \right) \left( \boldsymbol{y}_{H}^{ j } \right)\\
\end{array} \right], 
\end{aligned}
\label{eq:tripletb}
\end{equation}
here $t_n$ is the current time step, in which the settlement field reconstruction should be carried out. $N_{hp}$ is the number of sensors installed for monitoring in real projects.

\begin{figure}[!t]
	\centering
	\includegraphics[scale=0.79]{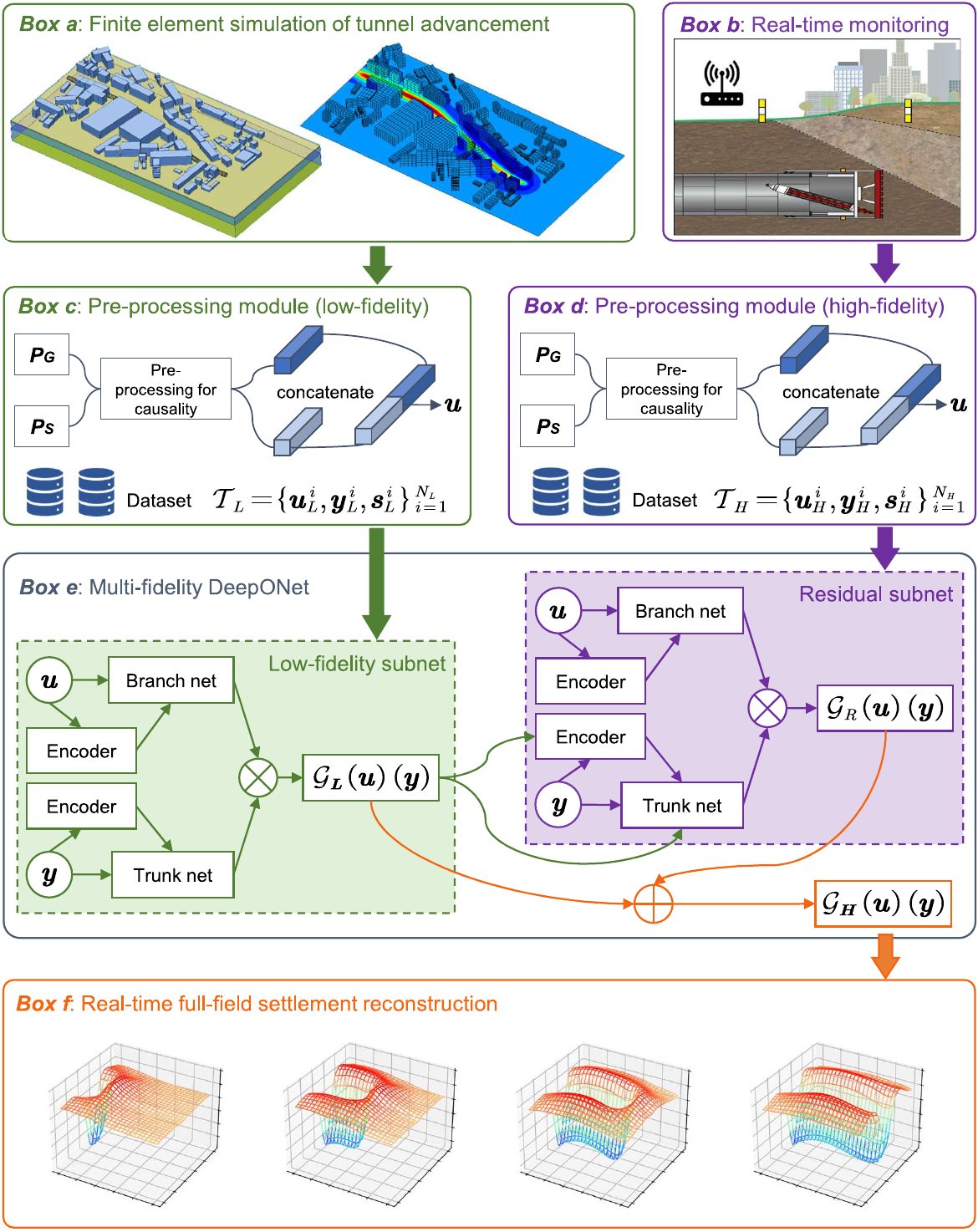}
	\caption{Illustration of key components of the developed multi-fidelity DeepONet with causality. The detailed description of the entire computational workflow in this figure can be referred to in Algorithm~\ref{alg:algo}.}
	\label{fig:MultiFidelity}
\end{figure}

In an effort to assimilate the aforementioned data sets $\mathcal{T}_L$ and $\mathcal{T}_H$, we present a multi-fidelity DeepONet architecture. As depicted in Box e of Fig.~\ref{fig:MultiFidelity}, the multi-fidelity DeepONet is composed of two sub-networks: a low-fidelity subnet and a residual subnet, both of which follow the modified DeepONet architecture presented in section~\ref{sec:ModiDON}. 

The low-fidelity subnet is trained using the low-fidelity data set $\mathcal{T}_L$ to learn the operator $\mathcal{G}_L$, which is capable of mapping arbitrary TBM process parameters to the corresponding ground settlement field. While the output of the low-fidelity subnet, $\mathcal{G}_L(\boldsymbol{u})(\boldsymbol{y})$, may not accurately fit the measured data from practical engineering projects, it can still capture the underlying physical trends of the real ground settlement field. The loss function for the low-fidelity subnet is formulated as:
\begin{equation}
\begin{aligned}
\mathcal{L}_{LF}&=\frac{1}{N_L}\sum_{i=1}^{N_L}{\begin{array}{c}
	\left( \mathcal{G}_L(\boldsymbol{u}_{L}^{i})(\boldsymbol{y}_{L}^{i})-\boldsymbol{s}_{L}^{i} \right)^2\\
\end{array}}
\\
&=\frac{1}{N_s\times N_t\times N_{lp}}\sum_{k=1}^{N_s}{\sum_{t_i=1}^{N_t}{\sum_{j=1}^{N_{lp}}{\left( \mathcal{G}_L(\boldsymbol{u}_{L,k}^{t_i})(\boldsymbol{y}_{L,k}^{j})-\boldsymbol{s}_{L,k}\left( \boldsymbol{u}_{L,k}^{t_i} \right) \left( \boldsymbol{y}_{L,k}^{j} \right) \right) ^2}}},
\end{aligned}
\label{eq:lossl}
\end{equation}
where $\boldsymbol{u}_{L,k}^{t_i}$, $\boldsymbol{y}_{L,k}^{j}$, and $\boldsymbol{s}_{L,k}\left( \boldsymbol{u}_{L,k}^{t_i} \right) \left( \boldsymbol{y}_{L,k}^{j} \right)$ actually correspond to their counterparts in Eq.~\eqref{eq:triplet}, and the newly introduced subscript $k$ is to indicate the specific number of the scenario.

The residual subnet is trained with the high-fidelity data set $\mathcal{T}_H$, and the learned operator $\mathcal{G}_R$ 
aims to characterize the complex correlations between low-fidelity and high-fidelity data. In this work, we employ the input augmentation approach proposed in Ref.~\citep{Lu_2022_multi-fidelity}, and the low-fidelity prediction $\mathcal{G}_L(u)(y)$ is appended to the trunk net inputs to make the residual operator $\mathcal{G}_R$ easier to learn. Hence, the final output of the multi-fidelity DeepONet, $\mathcal{G}_H(\boldsymbol{u})(\boldsymbol{y})$, is given by:
\begin{equation}
\begin{aligned}
\mathcal{G}_H(\boldsymbol{u}_H)(\boldsymbol{y}_H)=\mathcal{G}_L(\boldsymbol{u}_H)(\boldsymbol{y}_H)+\mathcal{G}_R\left( \boldsymbol{u}_H \right) \left( \mathcal{G}_L(\boldsymbol{u}_H)(\boldsymbol{y}_H),\boldsymbol{y}_H \right) .
\end{aligned}
\end{equation}

The loss function for training the residual subnet is expressed as:
\begin{equation}
\begin{aligned}
\mathcal{L}_{HF}&=\frac{1}{N_H}\sum_i^{N_H}{\left( \mathcal{G}_H(\boldsymbol{u}_{H}^{i})(\boldsymbol{y}_{H}^{i})-\boldsymbol{s}_{H}^{i} \right) ^2}
\\
&=\frac{1}{t_n\times N_{hp}}\sum_{t_i=1}^{t_n}{\sum_{j=1}^{N_{hp}}{\left( \mathcal{G}_H(\boldsymbol{u}_{H}^{t_i})(\boldsymbol{y}_{H}^{j})-\boldsymbol{s}_H\left( \boldsymbol{u}_{H}^{t_i} \right) \left( \boldsymbol{y}_{H}^{j} \right) \right) ^2}}.
\end{aligned}
\label{eq:lossh}
\end{equation}

The two subnets in the proposed multi-fidelity DeepONet are separately trained in two independent stages, differing from the simultaneous training approach presented in \citep{Lu_2022_multi-fidelity,Howard_2022_fidelityDON}. At the offline stage, the low-fidelity subnet is firstly trained using the data set $\mathcal{T}_L$ generated by finite element simulations, and then the trained network parameters are saved. At the online stage, the pre-trained low-fidelity subnet is loaded as a frozen module of the entire network, and only the residual subnet is trained with the data set $\mathcal{T}_H$ obtained from real-time monitoring in practical projects. Although the arrangement of monitoring points at the offline stage is normally sparse, the physics extracted from a large amount of low-fidelity data allows for reliable extrapolation of settlement values for points without monitoring, thereby enabling a comprehensive reconstruction of ground settlement across the field. The overall workflow is summarized in Algorithm~\ref{alg:algo}.

\begin{algorithm}[H]
\caption{A multi-fidelity deep operator network with causality for real-time
full-field settlement reconstruction in mechanized tunneling} \label{alg:algo}
\begin{algorithmic}[1]
\Statex \textbf{\underline{Offline stage}}
\State Build a finite element model representing the tunnel advancement process within predefined sections of the tunnel project.
\State Determine typical ranges for possible variations of the time series $\boldsymbol{P_G}$ and $\boldsymbol{P_S}$, with discrete time steps $t_i = 1,2, \cdots,N_t$.
\State Run simulations with $N_s$ different scenarios of $\boldsymbol{P_G}$ and $\boldsymbol{P_S}$ and record the numerical results of the ground settlement at $N_{lp}$ surface points. \Comment{Box (a) in Fig.~\ref{fig:MultiFidelity}}
\State Generate and pre-processing the low-fidelity data set $\mathcal{T}_L$ in the form of a triplet (see Eq.~\eqref{eq:triplet}) for causality. \Comment{Box (c) in Fig.~\ref{fig:MultiFidelity}}
\State Train the low-fidelity subnet with $\mathcal{T}_L$ by minimizing the loss function Eq.~\eqref{eq:lossl}.
\State Save the optimized low-fidelity subnet parameters $\theta_{L}$.
\Statex \textbf{\underline{Online stage (current time step $t_n$)}}
\State Obtain the recorded history of $\boldsymbol{P_G}$ and $\boldsymbol{P_S}$ from time step 1 to $t_n$.
\State Collect the ground settlement data at limited $N_{hp}$ monitoring points from time step 1 to $t_n$. \Comment{Box (b) in Fig.~\ref{fig:MultiFidelity}}
\State Generate and pre-processing the high-fidelity data set $\mathcal{T}_H$ in the form of a triplet (see Eq.~\eqref{eq:tripletb}) for causality. \Comment{Box (d) in Fig.~\ref{fig:MultiFidelity}}
\State Load the low-fidelity subnet parameters $\theta_{L}$ as a pre-trained model and set these parameters non-trainable.
\State Train the residual subnet with $\mathcal{T}_H$ by minimizing the loss function Eq.~\eqref{eq:lossh}.
\State Use the multi-fidelity DeepONet to extrapolate the settlement values at non-monitored points, thus reconstructing the whole field at the time step $t_n$. \Comment{Box (f) in Fig.~\ref{fig:MultiFidelity}}
\State Repeat steps 7-12 for the next step $t_{n+1}$.
\end{algorithmic}
\end{algorithm}

\section{Applications to ground settlement field reconstruction}
\label{sec:Resu}
\subsection{Overview}
\begin{figure}[!t]
	\centering
	\includegraphics[width=\textwidth]{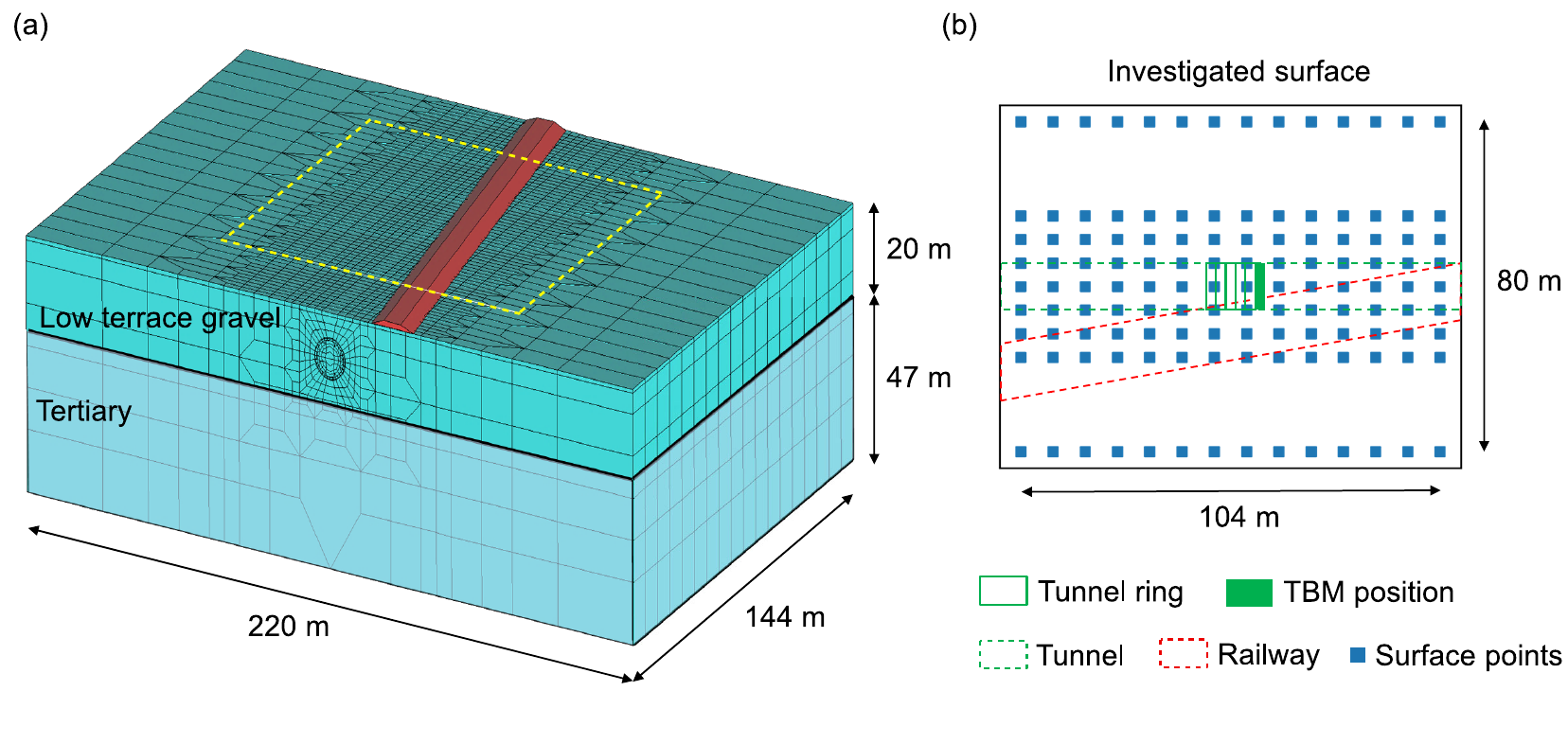}
	\caption{Numerical simulation model of a tunnel section: (a) geometry and FE discretization and (b) the investigated surface area.}
	\label{fig:ex_fem}
\end{figure}

We utilize the Data from a tunnel project in southern Germany to establish a numerical model, which is a real-world instance of the finite element framework discussed in section~\ref{sec:FEM}. Figure~\ref{fig:ex_fem} provides a visual representation of the numerical model, depicting a tunnel section constructed by a TBM. Within this tunnel section, two rail tracks are positioned on the top surface of a compacted ballast layer. The TBM is anticipated to cross beneath the railway, and the predicted settlement field can be useful for evaluating the structural safety of the above-ground rail tracks.

The dimensions of the model along the $x_1$, $x_2$, and $x_3$ directions, specifically 144~m, 220~m, and 67~m, respectively, have been thoughtfully determined to mitigate boundary effects. The lateral boundaries of the simulation domain are constrained in accordance with the normal directions of their respective surfaces, allowing vertical displacements while prohibiting horizontal movements. Meanwhile, the bottom boundary is fixed horizontally and vertically, following the standard practice in geotechnical applications of the FEM, owing to negligible deformation in deeper, high-stiffness soil layers.

The overburden (cover depth) and the tunnel diameter in this example are 6.6~m and 10.97~m, respectively. Each concrete tunnel lining ring measures 2~m in length and 0.5~m in thickness. The geological model, as depicted in Fig.~\ref{fig:ex_fem}(a), comprises two soil layers: a low terrace consisting of quaternary gravel (20~m thick) and tertiary material (47~m thick). The tunnel's excavation is entirely conducted within the top soil layer. Soil behavior is characterized by an elastoplastic model employing the Drucker-Prager yield criterion with linear isotropic hardening. The material properties of the lining ring, the shield machine, and the ballast layer are set to be linear elastic. The simulation encompasses different phases of tunnel advancement, including soil excavation, support pressure application, shield movement, grouting pressure application, and lining installation.

A rectangular area measuring 104~m in the $x_1$ direction and 80~m in the $x_2$ direction is regarded as the area of interest. This selection accounts for the fact that the settlement in the surrounding sections tends to approach zero. To represent this investigated surface area, a set of 126 nodes (i.e., $N_{lp}=126$) from the mesh grid, see Fig.~\ref{fig:ex_fem}(b), is chosen.

A total number of 100 FE simulations (i.e., $N_s=100$) corresponding to 100 combinations of input parameters is carried out. In each simulation, the input parameters include a scenario of time varying levels of the grouting pressure $P_G$ and a scenario of applying the face support pressure $P_S$. The values of $P_G$ and $P_S$ in each time step are generated according to practical suggestions in tunnel construction, which adopts pressure values within [120, 220] kPa and [100, 200] kPa, respectively. The total time steps in each simulation is 64, i.e., $N_t=64$. Finally, the data set $\mathcal{T}_L$ can be generated in the form described in Eq.~\eqref{eq:triplet}.

All the machine learning processes were implemented in JAX \citep{jax2018github} and trained in Google Colab on a NVIDIA T4 GPU. We use the Glorot normal scheme \citep{pmlr-v9-glorot10a} to initialize the weights of all the DeepONet networks and adopt hyperbolic tangent activation function (Tanh). The parameters were optimized using the Adam optimizer \citep{Kingma_2015_Adam}. The learning rate is set via the \textit{optimizers.exponential$\_$decay} module in JAX, which is parameterized by a list given as: (initial value, decay steps, decay rate). Details on the hyperparameter settings are shown in Table~\ref{tab:hyperparam}.

\begin{table}[!t]
\centering
\caption{Hyperparameters settings in the codes.}
\begin{tabular}{c|c}
\hline
Parameter                & Value                 \\ \hline
Low-fidelity (LF) subnet size & 3 layers, 100 neurons \\
Residual subnet size     & 2 layers, 100 neurons \\
Learning rate            & (1e-3, 1000, 0.9)       \\
Initialization           & Glorot normal       \\
Batch size for LF subnet      & 200,000            \\
Batch size for Residual net& Full batch       \\
Activation function      & Tanh                  \\ \hline             
\end{tabular}
\label{tab:hyperparam}
\end{table}

\subsection{Performance of The Low-fidelity Subnet}
\label{sec:offline}
The objective at offline stage is to train the simulation-based low-fidelity subnet for acquiring the operator $\mathcal{G}_L$, which represents the fundamental physical patterns of TBM-induced settlement fields. Unless noted, the inputs to DeepONets, namely $\boldsymbol{u}_L$, $\boldsymbol{y}_L$, $\boldsymbol{u}_H$, and $\boldsymbol{y}_H$, have to be scaled in the range of $(0, 1)$ by min-max normalization. The distribution of low-fidelity surface points for the trunk input $\boldsymbol{y}_L$ is illustrated in Fig.~\ref{fig:ex_fem}(b). 
Here, the data set $\mathcal{T}_L$ is partitioned into training, validation, and testing sets, comprising $80\%$, $10\%$, and $10\%$ of the total 100 scenarios, respectively. The training process reached convergence after 5000 iterations.

In order to evaluate the accuracy of the predicted results we consider the coefficient of determination, denoted $R^2$:
\begin{equation}
\begin{aligned}
R^2=1-\frac{\sum_i{\left( \boldsymbol{s}_i\left( \boldsymbol{u}_i \right) \left( \boldsymbol{y}_i \right) -\mathcal{G}\left( \boldsymbol{u}_i \right) \left( \boldsymbol{y}_i \right) \right)^2}}{\sum_i{\left( \boldsymbol{s}_i\left( \boldsymbol{u}_i \right) \left( \boldsymbol{y}_i \right) -\boldsymbol{\bar{s}} \right)^2}},
\end{aligned}
\end{equation}
where $\boldsymbol{\bar{s}}$ is the the mean of the true data $\boldsymbol{s}_i(\boldsymbol{u}_i)(\boldsymbol{y}_i)$.

\begin{figure}[!t]
	\centering
	\includegraphics[scale=1.0]{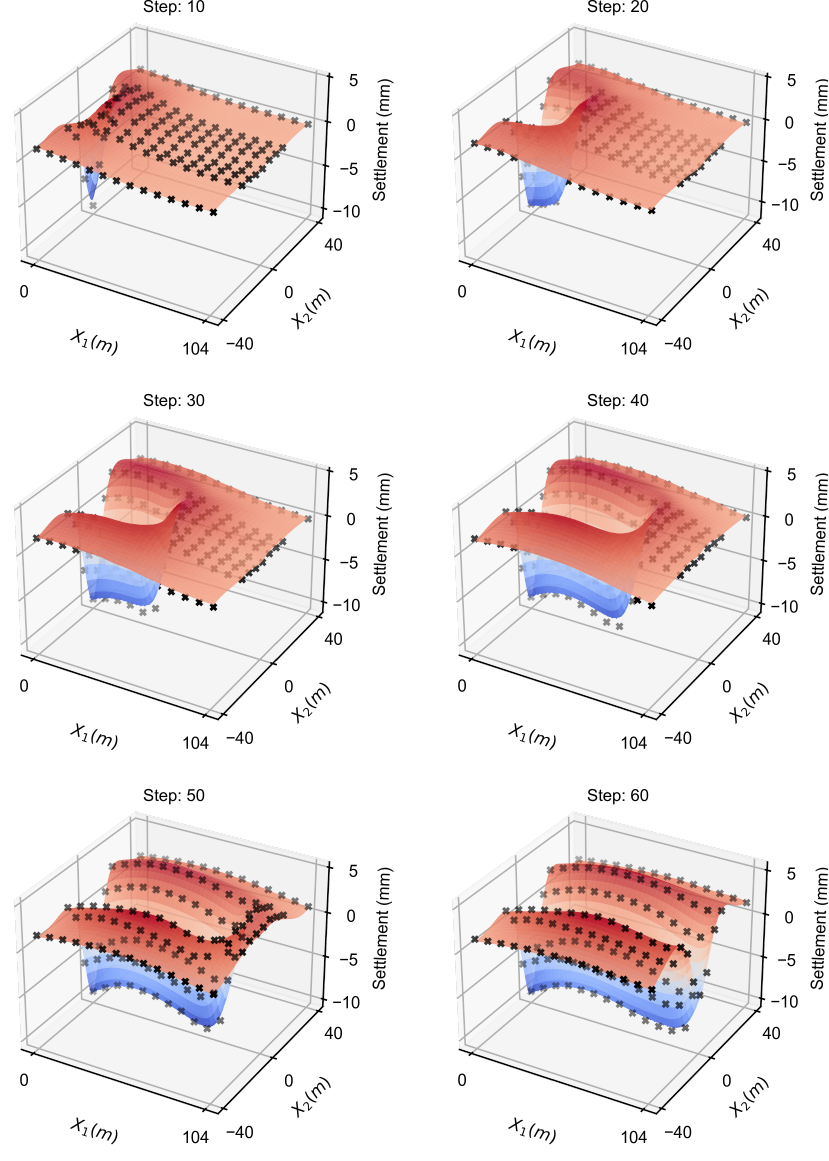}
	\caption{Snapshots of the settlement fields predicted by the low-fidelity subnet for the scenario $\#92$. True settlement data from FE simulations are marked with black crosses.}
	\label{fig:lf_results}
\end{figure}

For the testing set, the final $R^2$ score is 0.9867, indicating that the low-fidelity subnet is able to provide satisfactory predicted results with a similar accuracy compared to the physics-based simulations. The DeepONet can used as a reliable surrogate of finite element models for settlement predictions during mechanized tunneling. Figure~\ref{fig:lf_results} shows the evolution of the shape of settlement fields across different time steps in a selected scenario $\#92$. The pre-processing procedure for causality can effectively capture the temporal variation characteristics of the settlement field.

\subsection{Performance of The Proposed Multi-fidelity DeepONet}
\label{sec:online}
\begin{figure}[!t]
	\centering
	\includegraphics[scale=1.0]{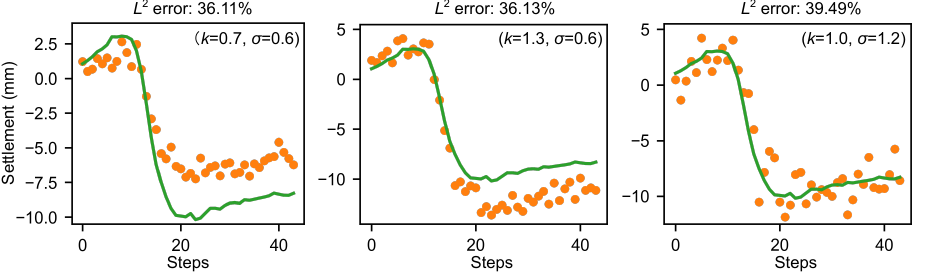}
	\caption{The mismatch between low-fidelity prediction (green curve, based on FE simulation data) and high-fidelity data (orange points, based on measurements) for the settlement evolution at a single point. The relationship between $k$ and 1.0 determines the type of error, i.e., "overestimated" ($k<1$), "underestimated" ($k>1$) and "fluctuated" ($k=1$).}
	\label{fig:errortype}
\end{figure}

\begin{figure}[!t]
	\centering
	\includegraphics[scale=1.0]{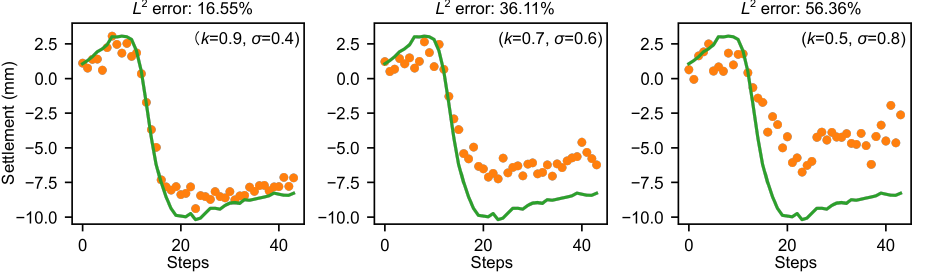}
	\caption{The mismatch between low-fidelity prediction (green curve, based on FE simulation data) and high-fidelity data (orange points, based on measurements) for the settlement evolution at a single point. When the type of error is fixed ($k<1$), different combinations of $k$ and $\sigma$ can lead to different error levels.}
	\label{fig:errorlevel}
\end{figure}

With the fact that a mismatch between simulation results and true measurement is obviously unavoidable in tunneling practice, in this paper, we alter the synthetic settlement data with several levels of modification and then employed the processed data as high-fidelity data to validate the effectiveness of our approach. 
In this manner, our proposed approach can prove the efficiency and the reliability in settlement prediction even in case of a fairly inaccurate prediction from the FE simulation.
In light of the environmental disturbances and sensor noise in engineering projects, the testing subset (including scenarios $\#91$ to $\#100$) from section~\ref{sec:offline} is selected and further modified to simulate the monitoring data in real. Specifically, the high-fidelity settlement data $\boldsymbol{s}_H$ can be generated by scaling low-fidelity settlement data $\boldsymbol{s}_L$ and adding some Gaussian noise, which reads:
\begin{equation}
\begin{aligned}
\underset{From\,\,measurement}{\underbrace{\boldsymbol{s}_H}}=\underset{From\,\,FE\,\,simulation}{\underbrace{\boldsymbol{s}_L}}\times k+\mathcal{N}\left( 0,\sigma ^2 \right) ,
\end{aligned}
\label{eq:lf2hf}
\end{equation}
where $k$ is the scaling factor and $\mathcal{N}\left( 0,\sigma ^2 \right)$ is the normal distribution with a mean of 0 and a standard deviation of $\sigma$.
We employ $L^2$ error to quantify the discrepancy between the obtained $\boldsymbol{s}_H$ and their original counterparts $\boldsymbol{s}_L$. Based on the exact value of $k$, the error can be categorized into three types: "overestimated" ($k<1$), "underestimated" ($k>1$) and "fluctuated" ($k=1$), as illustrated in Fig.~\ref{fig:errortype}. By adjusting $k$ and $\sigma$, we can generate synthetic monitoring data $\boldsymbol{s}_H$ with varying levels of error, which is shown in Fig.~\ref{fig:errorlevel}.
\begin{figure}[!t]
	\centering
	\includegraphics[scale=1.0]{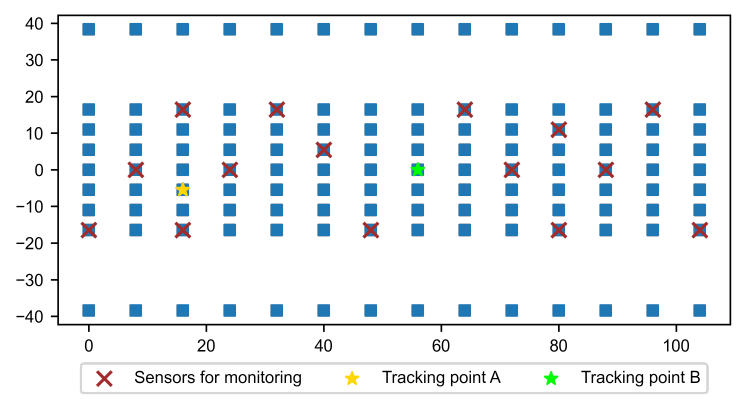}
	\caption{The layout of sensors for ground settlement monitoring. The blue square points are consistent with the surface points in Fig.~\ref{fig:ex_fem}(b). The tracking points here are used in sections \ref{sec:errortype} and \ref{sec:errorlevel} to record the evolution of ground settlement for individual points over time steps.}
	\label{fig:layout}
\end{figure}

The objective at the online stage is to learn the operator $\mathcal{G}_H$ from the measured data collected by a limited number of sensors, thus allowing real-time extrapolation of the settlement value for all points across the entire surface. Figure~\ref{fig:layout} displays the layout of the sensors used for settlement monitoring. A total of 15 grid points (i.e., $N_{hp}=15$) are selected from the fixed mesh grid in Fig.~\ref{fig:ex_fem}(b) for placing the sensors. According to Algorithm~\ref{alg:algo}, the low-fidelity subnet has been pre-trained during the offline stage, with only the residual subnet requiring further training at the online stage. Due to the limited volume of the high-fidelity data set and the small size of the residual subnet, the final training convergence for all cases mentioned in Table~\ref{tab:scenarios} can be achieved within only 200 iterations. By exploiting Jax's composable function transformations, the computing time for field reconstruction at each time step is only less than 1 minute. Without the transfer learning of the low-fidelity subnet, the online training of multi-fidelity DeepONet has to be trained from scratch, which takes approximately 30 minutes.

In the following three subsections, comprehensive examinations are carried out to evaluate the performance of the multi-fidelity DeepONet in the context of settlement prediction, as detailed in Table~\ref{tab:scenarios}. Section~\ref{sec:errortype} delves into the influence of various error types at equivalent magnitudes. Section~\ref{sec:errorlevel} assesses the efficacy of our methodology when confronted with errors of the identical type but varying magnitudes. Section~\ref{sec:minimum} discusses the minimum data volume required for settlement field reconstruction.
\begin{table}[!t]
\centering
\caption{Summary of case studies for assessing the performance of multi-fidelity DeepONets.}
\begin{tabular}{|c|c|c|c|c|c|c|}
\hline
Section                          & Scenario                 & Error type                & Error level                           & Time step                                       & Tracking point      & Results \\ \hline
                                 & \#91                    & \cellcolor[HTML]{CADFE8}$k<1$ &                                       &                                                 &                     &  Fig.~\ref{fig:scena91}      \\ \cline{2-3} \cline{7-7} 
                                 & \#95                    & \cellcolor[HTML]{CADFE8}$k>1$ &                                       &                                                 &                     &  Fig.~\ref{fig:scena95}      \\ \cline{2-3} \cline{7-7} 
\multirow{-3}{*}{\S\ref{sec:errortype}}  & \#99              & \cellcolor[HTML]{CADFE8}$k=1$ & \multirow{-3}{*}{$\approx30$\%}       & \multirow{-3}{*}{38}                            & \multirow{-3}{*}{A in Fig.~\ref{fig:layout}} &  Fig.~\ref{fig:scena99}      \\ \hline
                                 &                          &                           & \cellcolor[HTML]{CADFE8}$\approx15$\% &                                                 &                     &  Fig.~\ref{fig:scena100_10}      \\ \cline{4-4} \cline{7-7} 
                                 &                          &                           & \cellcolor[HTML]{CADFE8}$\approx35$\% &                                                 &                     &  Fig.~\ref{fig:scena100_30}      \\ \cline{4-4} \cline{7-7} 
\multirow{-3}{*}{\S\ref{sec:errorlevel}} & \multirow{-3}{*}{\#100} & \multirow{-3}{*}{$k<1$} & \cellcolor[HTML]{CADFE8}$\approx55$\% & \multirow{-3}{*}{54}                            & \multirow{-3}{*}{B in Fig.~\ref{fig:layout}} &  Fig.~\ref{fig:scena100_50}      \\ \hline
                                 & \#93                    & $k<1$                         &                                       & \cellcolor[HTML]{CADFE8}                        &                     &        \\ \cline{2-3} 
                                 & \#94                    & $k>1$                         &                                       & \cellcolor[HTML]{CADFE8}                        &                     &        \\ \cline{2-3} 
                                 & \#97                    & $k=1$                         & \multirow{-3}{*}{$\approx$30\%}       & \cellcolor[HTML]{CADFE8}                        &                     & \multirow{-3}{*}{Fig.~\ref{fig:minimum}(a)} \\ \cline{2-4} \cline{7-7} 
                                 & \#92                    & $k<1$                         &                                       & \cellcolor[HTML]{CADFE8}                        &                     &        \\ \cline{2-3} 
                                 & \#96                    & $k>1$                         &                                       & \cellcolor[HTML]{CADFE8}                        &                     &        \\ \cline{2-3} 
\multirow{-6}{*}{\S\ref{sec:minimum}} & \#98                    & $k=1$                         & \multirow{-3}{*}{$\approx$50\%}       & \multirow{-6}{*}{\cellcolor[HTML]{CADFE8}15 - 64} & \multirow{-6}{*}{-} &  \multirow{-3}{*}{Fig.~\ref{fig:minimum}(b)}      \\ \hline
\end{tabular}
\label{tab:scenarios}
\end{table}

\subsubsection{Investigations into The Effect of Error Type}
\label{sec:errortype}
\begin{figure}[!t]
	\centering
	\includegraphics[scale=1.0]{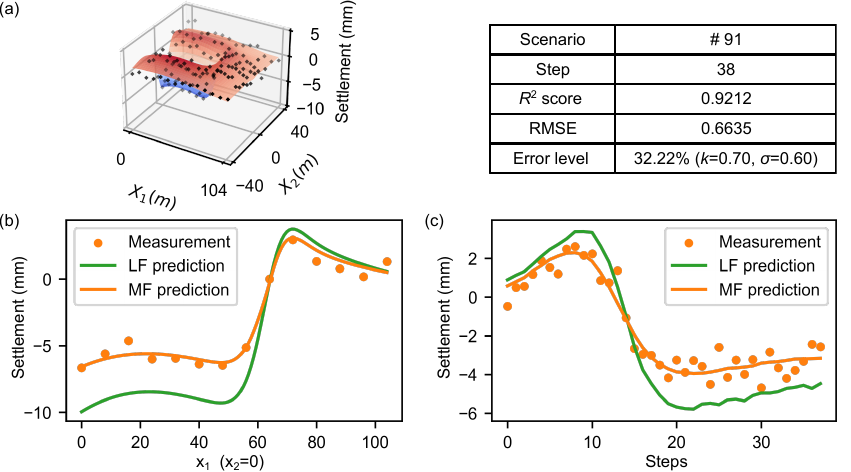}
	\caption{Predicted results in scenario $\#91$ (error type: $k<1$, error level: $\approx30\%$): (a) the reconstructed settlement field (black dots: true settlement values), (b) longitudinal profiles of surface settlement, and (c) settlement evolution at the tracking point A. Low-fidelity (LF) prediction is based on only FE simulation data, and multi-fidelity (MF) prediction is based on FE simulation and limited measurement data.}
	\label{fig:scena91}
\end{figure}

\begin{figure}[!t]
	\centering
	\includegraphics[scale=1.0]{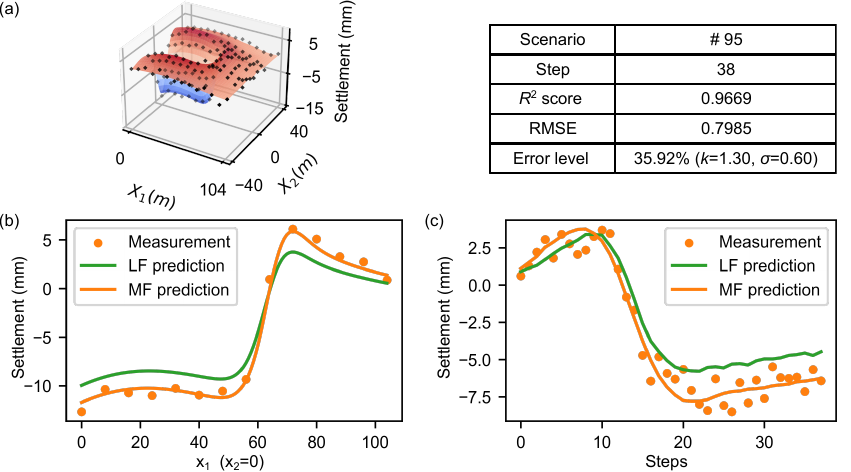}
	\caption{Predicted results in scenario $\#95$ (error type: $k>1$, error level: $\approx30\%$): (a) the reconstructed settlement field (black dots: true settlement values), (b) longitudinal profiles of surface settlement, and (c) settlement evolution at the tracking point A. Low-fidelity (LF) prediction is based on only FE simulation data, and multi-fidelity (MF) prediction is based on FE simulation and limited measurement data.}
	\label{fig:scena95}
\end{figure}

\begin{figure}[!t]
	\centering
	\includegraphics[scale=1.0]{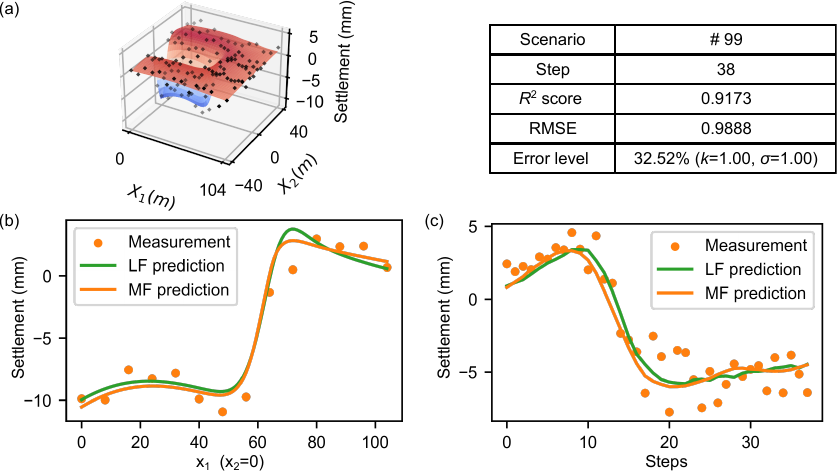}
	\caption{Predicted results in scenario $\#99$ (error type: $k=1$, error level: $\approx30\%$): (a) the reconstructed settlement field (black dots: true settlement values), (b) longitudinal profiles of surface settlement, and (c) settlement evolution at the tracking point A. Low-fidelity (LF) prediction is based on only FE simulation data, and multi-fidelity (MF) prediction is based on FE simulation and limited measurement data.}
	\label{fig:scena99}
\end{figure}

We randomly choose data from scenarios $\#91$, $\#95$, and $\#99$ and modify them via Eq.~\eqref{eq:lf2hf} to generate the measurement data. We set the value of $k$ value to $k<1$, $k>1$, and $k=1$ while maintaining the final error level around $30\%$. We showcase the effectiveness of our algorithm through settlement predictions at time step 38 for all three cases, as depicted in Figures \ref{fig:scena91}, \ref{fig:scena95}, and \ref{fig:scena99}. Each figure includes the reconstructed 3D settlement field, surface settlement curve at longitudinal section along the tunnel axis $x_2=0$, and the ground settlement evolution versus time step for a tracking point. 

In case of an "overestimated" scenario ($k<1$), the simulation predicts larger settlements, resulting in a mismatch of around 4~mm in the maximum settlement prediction within the longitudinal cross-section, see Fig.~\ref{fig:scena91}(b). However, the multi-fidelity (MF) DeepONet can still rectify the inaccuracy from the simulation-based low-fidelity (LF) subnet and ensure the accurate prediction of settlement values for other surface points along the cross section.
As shown in Fig.~\ref{fig:scena91}(c), the maximum settlement of a tracking point deviates by 2~mm compared to actual measurements. However, by incorporating data obtained from other surface points, the settlement of the tracking point is better predicted using the MF DeepONet.
Similar findings can also be observed in Figs~\ref{fig:scena95} and \ref{fig:scena99}, where other types of possible errors between simulation-based results and measurements are considered.
In general, the settlement field reconstruction results in all three cases yield $R^2$ scores above 0.9, indicating satisfactory accuracy can be obtained by our approach. 

It is worth noting that the multi-fidelity DeepONet output not only preserves the shape and trend of predictions from the low-fidelity model but also exhibits a satisfactory agreement with the high-fidelity measurement data. This achievement represents an effective fusion between physics-based simulations and real-world observations. Even with a limited number of monitoring sensors, our method is able to reconstruct the whole settlement field with support of the underlying physics embedded in finite element simulations.

\subsubsection{Investigations into The Effect of Error Level}
\label{sec:errorlevel}
\begin{figure}[!t]
	\centering
	\includegraphics[scale=1.0]{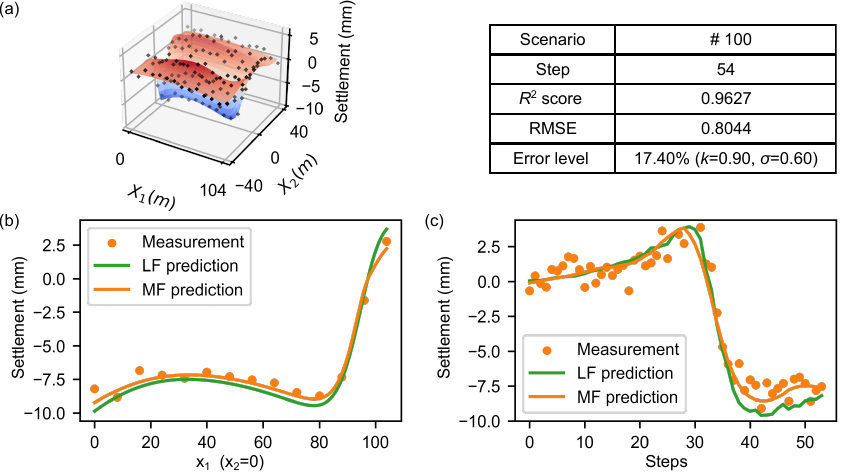}
	\caption{Predicted results in scenario $\#100$ (error type: $k<1$, error level: $\approx15\%$): (a) the reconstructed settlement field (black dots: true settlement values), (b) longitudinal profiles of surface settlement, and (c) settlement evolution at the tracking point A. Low-fidelity (LF) prediction is based on only FE simulation data, and multi-fidelity (MF) prediction is based on FE simulation and limited measurement data.}
	\label{fig:scena100_10}
\end{figure}

\begin{figure}[!t]
	\centering
	\includegraphics[scale=1.0]{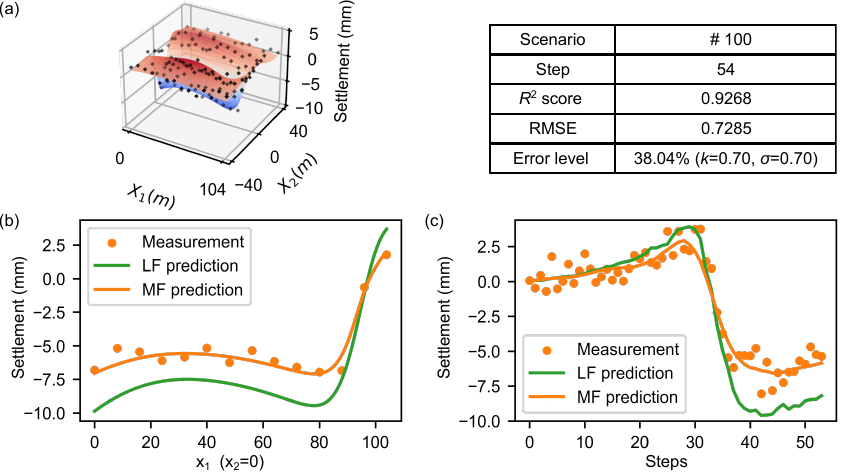}
	\caption{Predicted results in scenario $\#100$ (error type: $k<1$, error level: $\approx35\%$): (a) the reconstructed settlement field (black dots: true settlement values), (b) longitudinal profiles of surface settlement, and (c) settlement evolution at the tracking point A. Low-fidelity (LF) prediction is based on only FE simulation data, and multi-fidelity (MF) prediction is based on FE simulation and limited measurement data.}
	\label{fig:scena100_30}
\end{figure}

\begin{figure}[!t]
	\centering
	\includegraphics[scale=1.0]{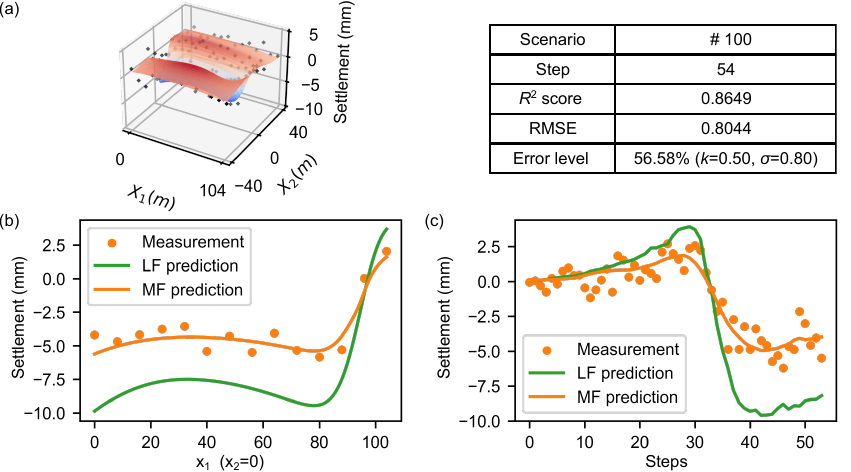}
	\caption{Predicted results in scenario $\#100$ (error type: $k<1$, error level: $\approx55\%$): (a) the reconstructed settlement field (black dots: true settlement values), (b) settlement curves along longitudinal-section $x_2=0.5$, and (c) settlement evolution at the tracking point A. Low-fidelity (LF) prediction is based on only FE simulation data, and multi-fidelity (MF) prediction is based on FE simulation and limited measurement data.}
	\label{fig:scena100_50}
\end{figure}

In this section, the data from scenario $\#100$ are selected for generating high-fidelity data. The value of $k$ in Eq.~\eqref{eq:lf2hf} is kept less than 1, which means herein we only focus on the "overestimated" ($k<1$) cases. After the modification to settlement data from FE simulations, measurement data with $L^2$ errors of $15\%$, $35\%$, and $55\%$ can be obtained for comparative study. The accuracy of the proposed algorithm is presented through settlement field reconstructions at time step 54, as shown in Figures \ref{fig:scena100_10}, \ref{fig:scena100_30}, and \ref{fig:scena100_50}.

When the error level between direct simulations (LF DeepONet) and measurement data is slight (approximately 15\%), see Fig.~\ref{fig:scena100_10}, the MF DeepONet certainly shows excellent agreements in the prediction of both settlement profile (Fig.~\ref{fig:scena100_10}(b)) and settlement evolution (Fig.~\ref{fig:scena100_10}(c)) with an overall $R^2$ value of 0.9627.
This performance metric exhibits a decrease to 0.9268 when the error level approaches around 35\% (Fig.~\ref{fig:scena100_30}).
In this case, the employment of the MF DeepONet results in a more accurate prediction of the maximum settlement, estimating it as 7.5~mm, as compared to the 10~mm settlement prediction when solely relying on simulation results (LF DeepONet).
In the most challenging case within our study, the error level between simulations and real measurements increases to roughly 55\% as shown in Fig.~\ref{fig:scena100_50}.
In this context, the MF DeepONet leverages both the imprecise information from simulations and limited measurement data to provide a reliable settlement prediction with an $R^2$ value of 0.8649. 
Despite the presence of considerable noise, the maximum settlement is predicted with higher precision, around 5~mm, compared to the 10~mm maximum settlement exclusively predicted by the LF DeepONet.

Generally speaking, it can be observed that as the error level increases, there is a corresponding decrease in the $R^2$ score for the predicted settlement field.  Nevertheless, even when the error level of measurements surpasses $50\%$, multi-fidelity DeepONet can still achieve prediction accuracy with an $R^2$ score exceeding 0.8, which is acceptable for practical engineering applications. With regard to the longitudinal profiles of surface settlement and the evolution of settlement at the tracking point, the results obtained in all three cases align well with measurements and follow similar physical patterns to that of numerical simulations.

It is widely recognized that training neural networks with extremely limited noisy data often leads to overfitting. However, in this work, the incorporation of physics through the low-fidelity subnet effectively mitigates such an overfitting problem, thus demonstrating strong robustness against noise.

\subsubsection{Investigations into the Effect of the Necessary Amount of Training Data}
\label{sec:minimum}
\begin{figure}[!t]
	\centering
	\includegraphics[scale=1.0]{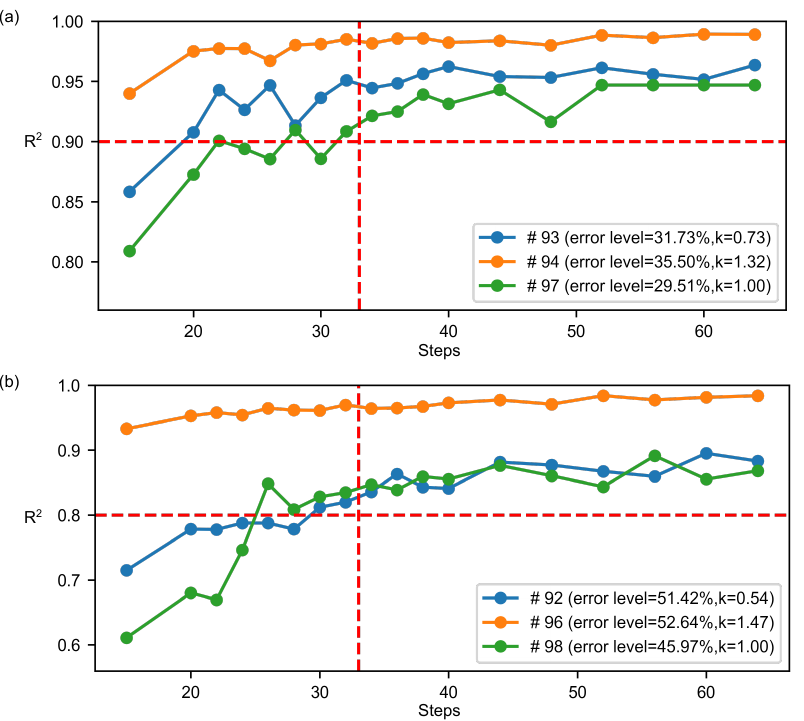}
	\caption{Accuracy of settlement field reconstructions at time steps 15-64 for all three error types at two different levels: (a) error level $\approx30\%$ and (b) error level $\approx50\%$.}
	\label{fig:minimum}
\end{figure}

Considering that in the initial excavation time steps, the available measured data for training might be insufficient to capture the correlation between high-fidelity and low-fidelity data; therefore, the minimum data volume required for settlement field reconstruction has to be studied in this section. Given that the number of monitoring points is fixed in this study, our attention is directed towards determining the threshold time step for prediction accuracy convergence. The analysis encompasses measurement data characterized by three different error types and two error levels, resulting in a total of six cases. Figure~\ref{fig:minimum} illustrates the $R^2$ scores of the settlement fields predicted by the multi-fidelity DeepONet across various time steps for these 6 cases.

It can be observed that in case of using a minimum number of 32 time steps for the network training (i.e., $t_n > 32$), 
$R^2$ scores exceeding 0.9 are achieved using measurements with a $30\%$ error level, while $R^2$ scores above 0.8 are obtained with measurements having a $50\%$ error level. 
By comparing Fig.~\ref{fig:minimum}(a) and Fig.~\ref{fig:minimum}(b), it can be seen that the predictive accuracy of our method tends to decrease in correspondence with increasing discrepancies between data obtained from finite element simulations and actual engineering measurements. It is noteworthy that results obtained in the case of "underestimated" ($k>1$) (i.e. orange lines in Fig.~\ref{fig:minimum}) significantly outperform those obtained from the other two error types. We speculate that this phenomenon may be attributed to the relatively large absolute values of measurements in this specific case. In contrast, the introduced Gaussian noise $\mathcal{N}\left( 0,\sigma ^2 \right)$ appears comparatively smaller, thereby yielding more precise predictions. 


\section{Conclusions}
\label{sec:Conc}
A multi-fidelity DeepONet framework has been adapted to perform real-time reconstruction of the settlement field induced by machine driven tunnel construction based on a priori trained data from computational simulation and data from continuous monitoring acquired during tunnel advancement at limited monitoring locations. 
The proposed DeepONet framework provides an efficient fusion of model based data, representing the physics of the problem with unavoidable simplifications and limited knowledge of parameters (low-fidelity data), and measurement data (high-fidelity data).
The main findings from this study can be summarized below: 
\begin{enumerate}[label=(\alph*)]
\item 
This reconstruction not only captures the anticipated trends derived from the physical model but also achieves satisfactory agreement with the observed data. 
\item The successful implementation of causality pre-processing can take into the account the time-dependent characteristics of both inputs and outputs. 
\item 
The incorporation of transfer learning significantly reduces training costs, leading to efficient field reconstruction for each excavation step in TBM operations. The computing time for every time step is less than one minute, showcasing great potential of our approach to be applied in tunnel engineering projects. 
\item 
The presented method exhibits robustness against noise, yielding $R^2$ scores above 0.8 even for small high-fidelity data sets with a $50\%$ error level. 
\end{enumerate}
Moreover, some potential improvements can be accomplished in future work: 
\begin{enumerate}[label=(\alph*)]
\item 
In this paper, the arrangement of monitoring points remains fixed throughout all analyses. Nevertheless, an upcoming study will extend the proposed multi-fidelity DeepONet framework to examine the influence of monitoring point layouts and suggest the optimal number and positions for monitoring sensors. 
\item 
Only two TBM process parameters, the tail void grouting pressure $\boldsymbol{P_G}$ and the face support pressure $\boldsymbol{P_S}$, are considered in this work. With the MIONet proposed in \citep{Jin2022}, the integration of more input parameters related e.g.  to a bandwidth of geotechnical conditions could be feasible. 
\item 
While the pre-processing procedure for causality can effectively deal with causal time series, it significantly inflates the size of the data set, which in turn influences training efficiency. A more powerful DeepONet architecture needs to be proposed to mitigate this challenge.
\end{enumerate}
It is emphasized, that although the technology proposed in this paper has specifically been applied to tunneling engineering, it can be used for the process control in a variety of engineering applications.

\section{Acknowledgment}
The first author acknowledges the support from the China Scholarship Council (CSC) under grant 202006260038. The authors also acknowledge financial support by the National Key Research and Development Program of China (Grant No. 2021YFE0114100), and the Science and Technology Commission of Shanghai Municipality (grant No. 22DZ1203005).



\clearpage
\bibliographystyle{apalike}
\bibliography{Reference}







\end{document}